\documentstyle[sprocl,epsf]{article}

\newcommand{\Tr}{\makebox{ Tr }}

\newcommand{\GeV}{\makebox{ GeV}}

\newcommand{\fm}{\makebox{ fm}}

\newcommand{\beq}{\begin{equation}}

\newcommand{\enq}{\end{equation}}

\newcommand{\beqa}{\begin{eqnarray}}

\newcommand{\enqa}{\end{eqnarray}}

\newcommand{\nn}{\nonumber}

\newcommand{\lbq}[1]{\label{#1} \enq}

\newcommand{\lbqa}[1]{\label{#1} \enqa}

\newcommand{\befi}[1]{\begin{figure}[ht] \leavevmode \centering \epsffile{#1.eps}}

\newcommand{\lbfi}[1]{\label{#1} \end{figure}}

\newcommand{\eq}[1]{eq.(\ref{#1})}

\newcommand{\fig}[1]{fig.(\ref{#1})}

\newcommand{\lbcap}[3]{\begin{minipage}{#1}\caption{\small #2}\label{#3}\end{minipage}\end{figure}}

\newcommand{\pa}{\partial}

\newcommand{\cD}{\mbox{$\cal D$}}

\newcommand{\cP}{\mbox{$\cal P$}}

\newcommand{\cN}{\mbox{$\cal N$}}

\newcommand{\bA}{\mbox{\bf A}}

\newcommand{\bF}{\mbox{\bf F}}

\newcommand{\al}{\alpha}

\newcommand{\be}{\beta}

\begin{document}
\title{LOW ENERGY THEOREMS AND THE SU(3)-FLUX-TUBE}
\author{\large Michael Rueter}
\address{Institut f\"ur theoretische Physik\\
Universit\"at Heidelberg\\
Philosophenweg 16, D-69120 Heidelberg, FRG\\
e-mail: M.Rueter@thphys.uni-heidelberg.de\\
\rm supported by the Deutsche Forschungsgemeinschaft
}
\maketitle\abstracts{The total energy and action, which is stored in the flux-tube between a static quark-antiquark pair, can be compared with the potential of the pair with the help of two low energy theorems. The flux-tube and the potential are calculated in the framework of the model of the stochastic vacuum. Using the low energy theorems we obtain consistency of the results and can predict the scale where the model describes the non-perturbative gluon dynamics of QCD.}
\section{Introduction}
The treatment of QCD processes involving large distances is very complicated, because perturbation theory is not applicable. The model of the stochastic vacuum (MSV) makes the crucial assumption that the complicated infra-red measure of the functional integral in non-Abelian gauge theories can be adequately approximated by a stochastic process with a converging cluster expansion [1]-[3]. This assumption already leads to a confining potential. In order to make quantitative predictions one has to make an additional strong simplification and assume the stochastic process to be a Gaussian one. For a detailed description and applications of the MSV we refer to [3][4].\\
In this note we work with an Euclidean field theory where the square of the electric field has the opposite sign of the same quantity
in a Minkowski field theory (MFT), whereas the square of the magnetic field has
the same sign in both theories:
\[  \vec E^2_{\rm EFT} = - \vec E^2_{\rm MFT}, \quad
  \vec B^2_{\rm EFT} =  \vec B^2_{\rm MFT}\; .
\]
Here and in the following a sum over the $N_C^2-1$ color components is understood.
\section{The low energy theorems}
By differentiating the expectation value of a Wilson-loop
\beqa
<W[L]> &=& \frac{1} {\cN} \int \cD A \exp[-S_{QCD}]  \Tr \cP \exp[-i\int_L
\bar{ \bA}_\mu d x_\mu],\nn  \\
\cN&\equiv &\int \cD A \exp[-S_{QCD}],\nn\\
S_{QCD}&\equiv&  \frac{1}{8 \pi \al_{s0}}\int {\rm d^4}x\Tr [\bar {\bF}_{\mu \nu}(x)\bar{ \bF}_{\mu \nu}(x)],\;\bar {\bF}_{\mu \nu} \equiv g_0 \bF^{(0)}_{\mu \nu}
\lbqa{3}
with respect to the unrenormalized coupling we obtained in [5] for the renormalized quantities:
\beq
 -\al_{s}\frac{\pa V(R)}{\pa \al_{s}} = \frac{1}{2}< \int {\rm d^3}x\Tr [\bF_{\mu \nu}(x)\bF_{\mu\nu}(x)]>_R,\lbq{7a}
where
\beq V(R) \equiv \lim_{T\to \infty}- \frac{\log <W[L]>}{T} \lbq{6}
and $< \; . \;  >_R$ shall denote the expectation value in the presence of a static quark-antiquark pair at distance $R$ where the expectation value in absence of the sources has been subtracted. By using renormalization group arguments we find
\beq 
\left\{ V(R) + R \frac{\pa V(R)}{\pa R} \right\}=\frac{1}{2}\frac{\tilde{\be}}{\al_{s}}< \int {\rm d^3}x\left(\vec{E}(x)^2+\vec{B}(x)^2 \right)>_R
\lbq{11}
with $\tilde{\be}(\al_{s}) \equiv \mu \frac{\rm d}{\rm d\mu} \al_{s}(\mu)$. This is the low energy theorem for the action and the one for the energy is the usual relation between the potential and the energy density:
\beq
V(R) = \frac{1}{2}< \int {\rm d^3} x \left( -\vec E(x)^2 +
\vec B(x)^2 \right)>_R.
\lbq{12}
For the case of a linear potential these are just two of the known low energy theorems [6]. In lattice QCD these theorems are known as Michael's sum rules [7], but there in the action sum rule scaling of $V(R)$ with $R$ was not taken into account.
\section{Consistency of the SU(3)-flux-tube with the low energy theorems}
\begin{figure}[ht]
\begin{minipage}{5.2cm}
\epsfysize4cm
\epsfxsize5.2cm
\epsffile{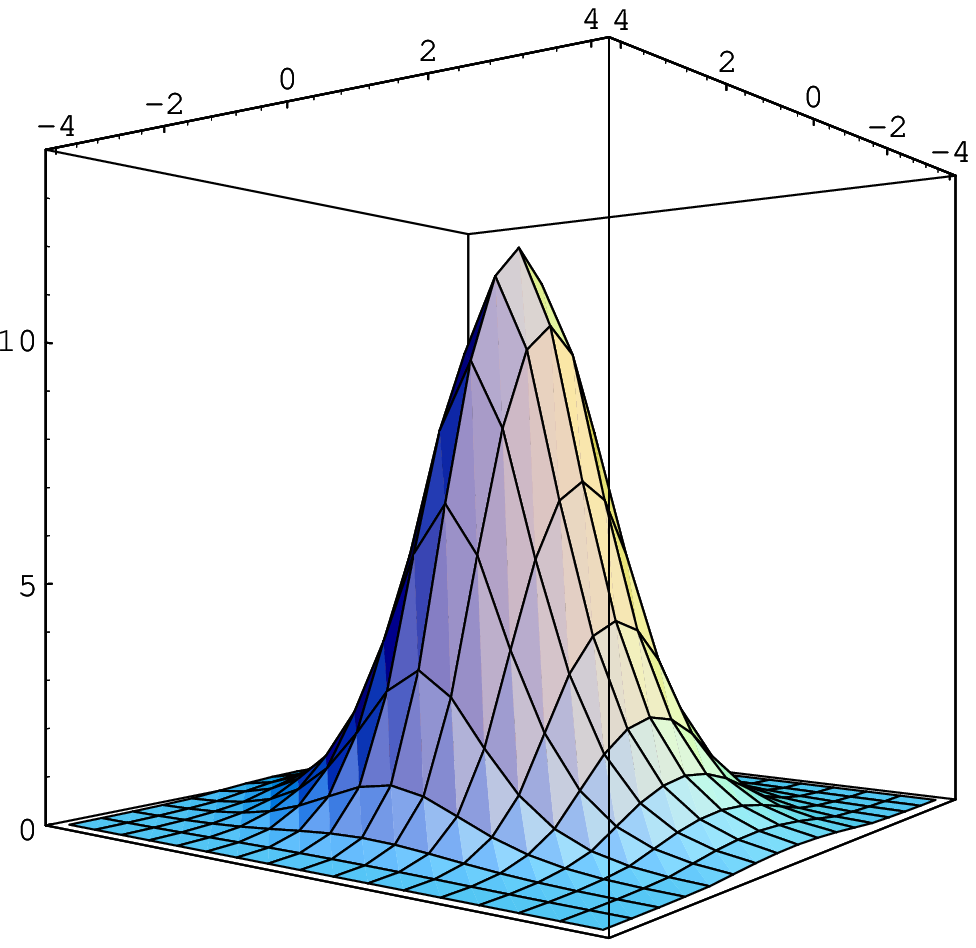}
$R=4a$, $a=$0.35fm
\end{minipage}
\hfill
\begin{minipage}{5.2cm}
\epsfysize4cm
\epsfxsize5.2cm
\epsffile{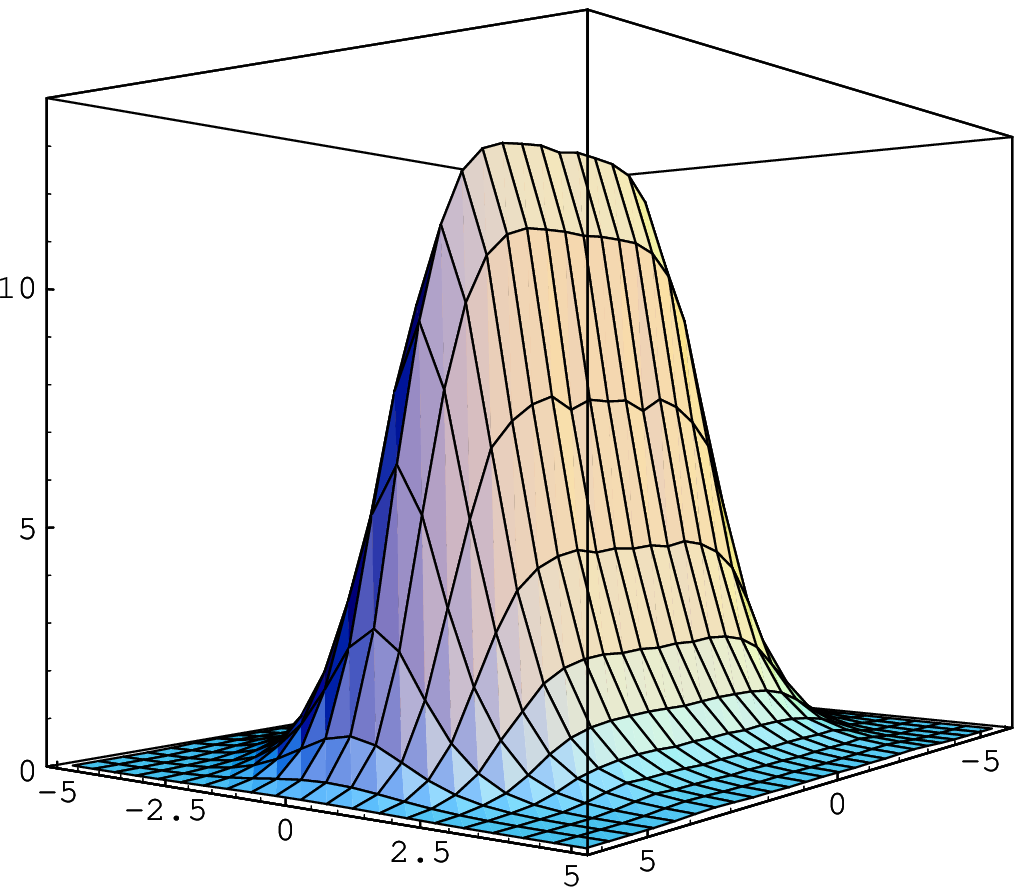}
$R=9a$, $a=$0.35fm
\end{minipage}
\caption{Difference of the squared electric field parallel to the quark-antiquark ($x_3$) axis \mbox{($-<g^2E_\|^2(x_3,x_\perp )>_R$)} in $\rm \frac{GeV}{fm^3}$ for different quark separations $R$. $x_\perp$ is the distance transverse to the $x_3$-axis and the dots denote the quark positions.}
\label{34c}
\unitlength1cm
\begin{picture}(0,0)
\put(.7,2.2){$x_\perp [a]$}
\put(7.3,2.2){$x_\perp [a]$}
\put(4.7,2.2){$x_3 [a]$}
\put(11.2,2.2){$x_3 [a]$}
\put(2.2,3.6){\circle*{.2}}
\put(8.7,3.5){\circle*{.2}}
\end{picture}
\end{figure}

\begin{figure}[ht]
\begin{minipage}[t]{5.5cm}
\centering
\epsfxsize5.5cm
\epsffile{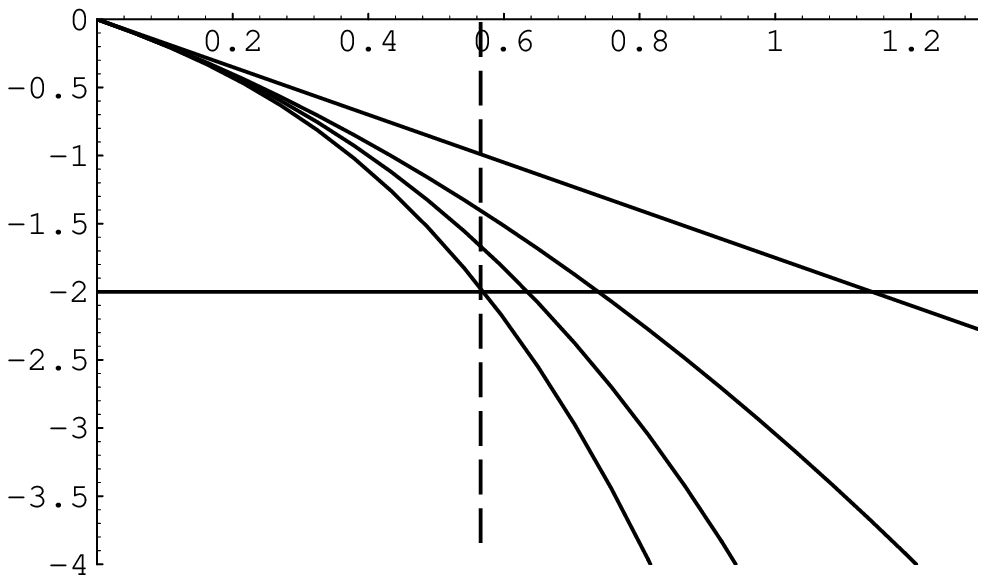}
\unitlength.8cm
\begin{picture}(0,0)
\put(-2.5,1.0){\Large $\frac{\tilde{\be}}{\al_s}$}
\put(2.0,4.1){$\al_s$}
\put(2,3.3){\scriptsize 1-loop}
\put(2.5,1.5){\scriptsize 2-loop}
\put(1.3,0.5){\scriptsize 3-loop $\overline{MS}$}
\put(-.7,.8){\scriptsize $\be_2=6250$}
\put(-.7,4.9){\scriptsize $\al_s=0.57$}
\end{picture}
\caption{$\frac{\tilde{\be}}{\al_s}$ as a function of $\al_s$ for the $\tilde{\be}$-function calculated in different loop order. The 3-loop results depends on the renormalization scheme. To fulfill $\tilde{\be}/\al_s$=-2 at $\al_s=0.57$ we fitted the 3-loop coefficient to be $\be_2=6250$, which agrees curiously with some lattice \mbox{results [9].}}
\label{btildeOVERalpha}
\end{minipage}
\hfill
\begin{minipage}[t]{5.5cm}
\centering
\epsfxsize5.5cm
\epsffile{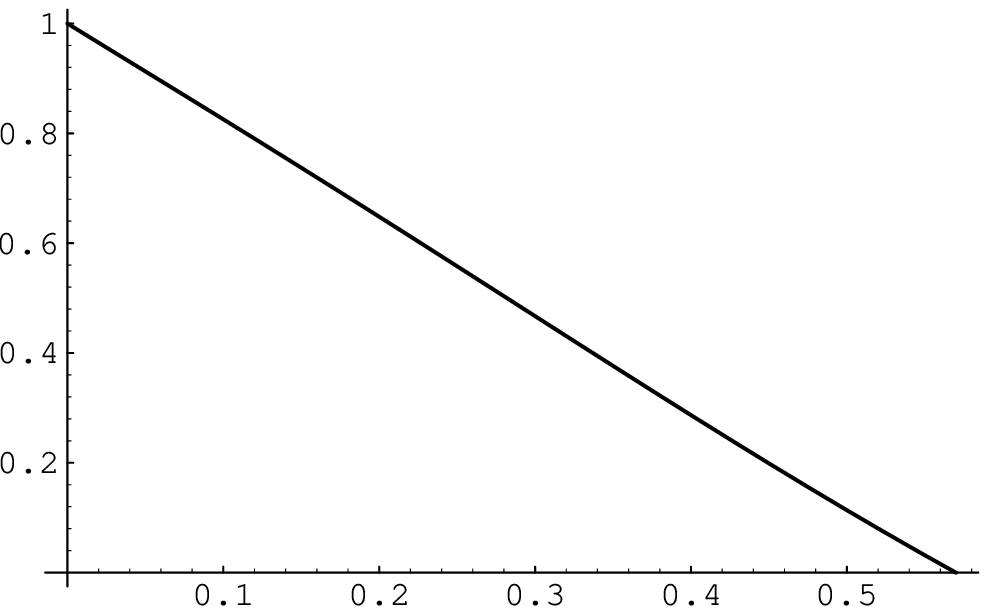}
\unitlength.8cm
\begin{picture}(0,0)
\put(3,0.2){$\al_s$}
\put(-3.8,4.5){$Q$}
\end{picture}
\caption{The ratio $Q$ as a function of $\al_s$}
\label{BOVERE}
\end{minipage}
\end{figure}

In ref.[8] we calculated the color-fields in the flux-tube between a static quark-antiquark pair. It was shown that by symmetry arguments $<g^2\vec{B}^2>_R=0$. The square of the electric field perpendicular to the loop $L$ ($<g^2E_\perp ^2>$) is also practically not influenced but only the squared electric field parallel to $L$ ($<g^2 E_\| ^2>$) is affected by the static sources. In \fig{34c} we display $-<g^2 E_\| ^2>_R$ as a function of the perpendicular distance $x_\perp$ from the loop $L$ and of the position along the quark-antiquark axis $x_3$. The lengths are given in units of the correlation length $a$ of the gluon field strengths, which is fixed to $a=0.35\fm$ in the MSV.
\parindent0em
By using the energy sum rule \eq{12} with $<g^2\vec{B}^2>_R=0$ we fix $\al_s=0.57$ at the largest $q\bar{q}$ distance. The comparison of the total energy stored in the flux-tube with the potential in then shown in \fig{Egesamt}.
\epsfxsize5.2cm
\befi{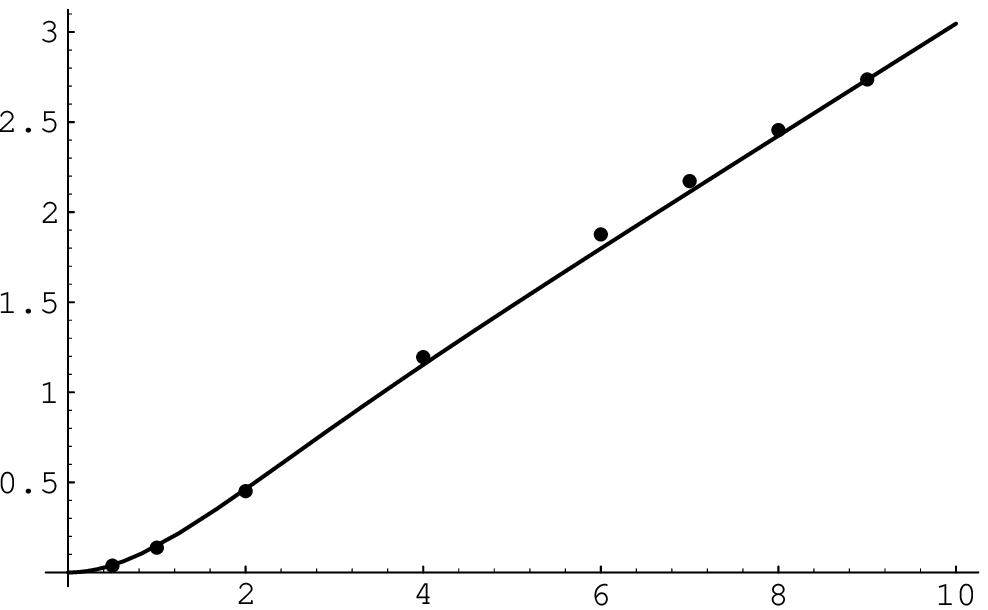}
\unitlength0.75cm
\begin{picture}(0,0)
\put(-1,.7){$R[a]$}
\put(-6.3,4){$V[\GeV ]$}
\end{picture}
\caption{The total energy stored in the field (dots) calculated with the energy sum rule (\protect\ref{12}) as compared with the potential of a q-$\rm \bar{q}$-pair obtained by evaluation of the Wilson-loop in the MSV [1][2] (solid line).}
\label{Egesamt}
\end{figure}\parindent0em
For a linear potential and $<g^2\vec{B}^2>_R=0$ \eq{11} and \eq{12} can only be fulfilled if $\tilde{\be}/\al_s$=-2 at the scale where $\al_s=0.57$. Using the perturbative calculated $\tilde{\be}$-function we check this condition in \fig{btildeOVERalpha}.
Finally we calculate the ratio of
\[Q\equiv \frac{\int {\rm d^3}x <\vec B(x)^2>_R}{\int {\rm d^3}x<\vec E(x)^2>_R}= \frac{2+\tilde{\be}/\al_s}{2-\tilde{\be}/\al_s} \]
which can be calculated in lattice gauge calculations.
\section*{Acknowledgments}
The author would like to thank H.G.~Dosch and O.~Nachtmann, with whom this work was done.


\section*{References}
\begin{thebibliography}{12}
\bibitem{1} H.G.~Dosch, {\it Phys.Lett.}{\bf B190} (1987) 177
\bibitem{2} H.G.~Dosch, Y.A.~Simonov, {\it Phys.Lett.}{\bf B205} (1988) 339
\bibitem{3} H.G.~Dosch, {\it Prog.Part.Nucl.Phys.}{\bf 33} (1994) 121
\bibitem{4} H.G.~Dosch, E.~Ferreira, A.~Kr\"amer, {\it Phys.Rev.}{\bf D50} (1994) 1992
\bibitem{5} H.G.~Dosch, O.~Nachtmann, M.~Rueter, {\it hepph-9503386}
\bibitem{6} V.A.Novikov, M.A.Shifman, A.I.Vainshtein, V.I.Zakharov, \\
{\it Nucl.Phys.}{\bf B191} (1981) 301\\
A.I.Vainshtein, V.I.Zakharov, V.A.Novikov, M.A.Shifman,\\
{\it Sov.J.Part.Nucl.}{\bf 13} (1982) 224
\bibitem{7} C.Michael, {\it Nucl.Phys.}{\bf B280} (1987) 13
\bibitem{8} M.Rueter, H.G.Dosch, {\it Z.Phys.}{\bf C66} (1995) 245 
\bibitem{9} M.L\"uscher, R.Sommer, P.Weisz, U.Wolff, {\it Nucl.Phys.}{\bf B413} (1994) 481
\end{thebibliography}
\end{document}